\documentclass[aps,pre,preprint,groupedaddress]{revtex4-1}

\usepackage{amsmath,amsfonts,amsthm} 
\usepackage{graphicx}
\usepackage{xcolor}

\newcommand{\be}{\begin{equation}}
\newcommand{\ee}{\end{equation}}
\newcommand{\ba}{\begin{eqnarray}}
\newcommand{\ea}{\end{eqnarray}}

\usepackage{epsfig}
\usepackage{float}

\usepackage{epsfig}
\usepackage{float}

\begin{document}

\title[Critical point scaling derivation]{Thermodynamic derivation of scaling at the liquid-vapor critical point}
\author{J.C. Obeso-Jureidini, D. Olascoaga  and V. Romero-Roch\'in}
\affiliation{Instituto de F\'isica, Universidad Nacional Aut\'onoma de M\'exico \\
Apartado Postal 20-364, 01000 Cd. M\'exico, Mexico}

\date{\today}

\begin{abstract}

With the use of thermodynamics and general equilibrium conditions only, we study the entropy of a fluid in the vicinity of the critical point of the liquid-vapor phase transition. By assuming a general form for the coexistence curve in the vicinity of the critical point, we show that the functional dependence of the entropy as a function of energy and particle densities necessarily obeys the scaling form hypothesized by Widom. Our analysis allows for a discussion on the properties of the corresponding scaling function, with the interesting prediction that the critical isotherm has the same functional dependence, between the energy and particles densities, as the coexistence curve. In addition to the derivation of the expected equalities of the critical exponents, the conditions that lead to scaling also imply that while the specific heat at constant volume can diverge at the critical point, the isothermal compressibility must do so. 
 
\end{abstract}

\maketitle

\section{Introduction}

The full thermodynamic description of critical phenomena in the liquid-vapor phase transition of pure substances has remained as a theoretical challenge for a long time \cite{Domb}. Substantially, the scaling hypothesis introduced by Widom \cite{Widom1965} proved to be a fundamental step in the understanding of experiments \cite{Michaels,Heller,Stewart,Mahajan} and numerical simulations \cite{Panagiotopoulos,Allen,Harris,Binder} of fluids in the vicinity of the critical point. This hypothesis establishes that if the free energies have a specific functional dependence on their state variables, say Helmholtz free energy in terms of particle density and  temperature, the critical exponents are not independent of each other obeying certain equalities \cite{Kadanoff,Fisher-review,review-SH}. These exponents characterize the behavior of the thermodynamic properties in the neighborhood of the critical point. Although a consequence of the equalities is that there are two independent exponents only, thermodynamics alone, being an empirical discipline, is unable to predict their numerical values. Indeed, the development of the renormalization group (RG) \cite{Wilson-Kogut,Ma,Amit} led to both, a validation of the scaling hypothesis and to a procedure to calculate the exponents in a systematic expansion involving the dimensionality of space. RG in turn is based on certain hypotheses regarding the partition functions of statistical mechanics, mainly scaling invariance close to the critical point. The transcendence of RG,  not only in the study of critical phenomena but in many other disciplines, cannot be exaggerated yielding a completely novel approach and understanding to the physics involved. Additionally, one of the major accomplishments of RG concerns the concept of {\it universality} that indicates that the critical exponents are the same not only for all chemically pure fluids but also for all solids showing ferromagnetism, in particular. Perhaps due to these successes the scaling {\it hypothesis} remained as such from a pure thermodynamic point of view, leaving the impression that thermodynamics alone, with its assumptions based on empirical observations, is truly unable to account for it. From this perspective, the purpose of this article is to show that the scaling hypothesis for the liquid-vapor phase transition can certainly be deduced using thermodynamics only. The present development generalizes the derivation of the scaling hypothesis for the para-ferromagnetic transition presented in Ref. \cite{yo}. As it can be contrasted, the difference between the derivation for a magnetic system, given in such a reference, with the present one for a liquid-vapor transition, is the lack of intrinsic symmetries of the latter, naturally included in the former. Although these results cannot show that the critical exponents have the same values for those two physically disimilar systems, the whole procedure is based after all in the laws of thermodynamics and the phase-equilibrium conditions which are universal for all substances in nature.\\

The derivation of scaling starts by analyzing the structure of the entropy per unit of volume $s$ as a function of the particle density $n$ and the internal energy per unit of volume $e$, namely, $s = s(e,n)$, visualized as a surface on a cartesian $(e,n,s)$ set of axes. The functional dependence of $s$ on $e$ and $n$ is fundamental in the sense that all the equilibrium thermodynamics properties of a pure fluid can be derived from it \cite{LLI,Callen}. The laws of thermodynamics indicate that $s$ is a concave single-valued function of $e$ and $n$ and that the intensive conjugate variables, temperature $T$ and chemical potential $\mu$, are continuous everywhere. Therefore, the empirical observation of the existence of a liquid-vapor first-order phase transition ending at a critical point, requires that the surface $(e,n,s)$ has a ``cut'' or void region, such that $e$, $n$ and $s$ are discontinuous at its edge but $T$ and $\mu$ continuous for all pairs of liquid and vapor coexisting states. The edge of such a void region is the {\it coexistence curve} of the transition. The critical point is identified solely as the ending point of the coexistence liquid and vapor states and we make absolutely no additional assumptions about it. As it will be specified, the shape of the surface and the coexistence curve can be quite complicated and, in principle, arbitrary in the $(e,n,s)$ axes, with no prescribed symmetries. However, by changing to a local set of coordinates with the origin at the critical point and along the principal axes of the surface, one can then argue that the coexistence curve is symmetric along the axis tangential to the critical point on the coexistence curve. This assumption is based on the fact that experimental and computer simulated coexistence curves are symmetric very near the critical point \cite{Heller,Stewart,Mahajan,Panagiotopoulos,Allen,Harris,Binder}; also, not shown here, it is an exercise to verify that the van der Waals model of a fluid \cite{LLI,Callen} also shows this symmetry. Concretely, the purpose of this article is to show that these very general considerations on the entropy surface and on the coexistence curve lead to the scaling hypothesis. That is, we show that the functional form of $s$ on $e$ and $n$, in the vicinity of the critical point, necessarily has the dependence hypothesized by Widom \cite{Widom1965}. A very important but natural consequence of the present analysis is that, while the specific heat at constant volume may or may not diverge at the critical point, the isothermal compressibility necessarily does diverge. We recall that the latter result is equivalent to the appearance of the unbounded density fluctuations and of the losing of all length scales at the critical point, which are the essence of RG\cite{Wilson-Kogut,Ma,Amit}. It is thus very exciting to find out that thermodynamics predicts this divergent behavior without appealing to the molecular structure of the fluid. A geometrically equivalent way to express these critical divergences is that the existence of a coexisting curve, bounding a void region on an otherwise concave function, implies the vanishing of the gaussian curvature of the surface at the critical point; that is, the surface forcibly becomes locally flat at such a point.\\

In Section 2 we present a brief summary of the general thermodynamic properties of the function $s = s(e,n)$. Section 3 is devoted to the isometrical transformation from the natural axes $(e,n,s)$ to an appropriate local set of coordinates at the critical point, in which one the axes is the normal to the surface, other is the tangent to the coexistence curve at the critical point, with the third one being orthogonal to the previous ones. In Section 4 we analyze the strong requirements that the coexistence curve imposes on the local entropy function and its derivatives, and show that these conditions straightforwardly imply scaling of the entropy function. We also discuss the general properties of the obtained scaling functions of the entropy. Section 5 is dedicated to the derivation of the usual critical exponents for the behavior of the density and chemical potential in terms of the temperature, as well as for the specific heat and constant volume and the isothermal compressibility, near the critical point. We conclude with some final remarks that we consider to be relevant. Details of some lengthy calculations and a generalization of the derivation of the scaling forms are given in two appendices.

\section{Thermodynamic conditions for the liquid-vapor phase transition in a pure fluid}

\begin{figure}[htbp]
\begin{center}
\includegraphics[width=.45\textwidth]{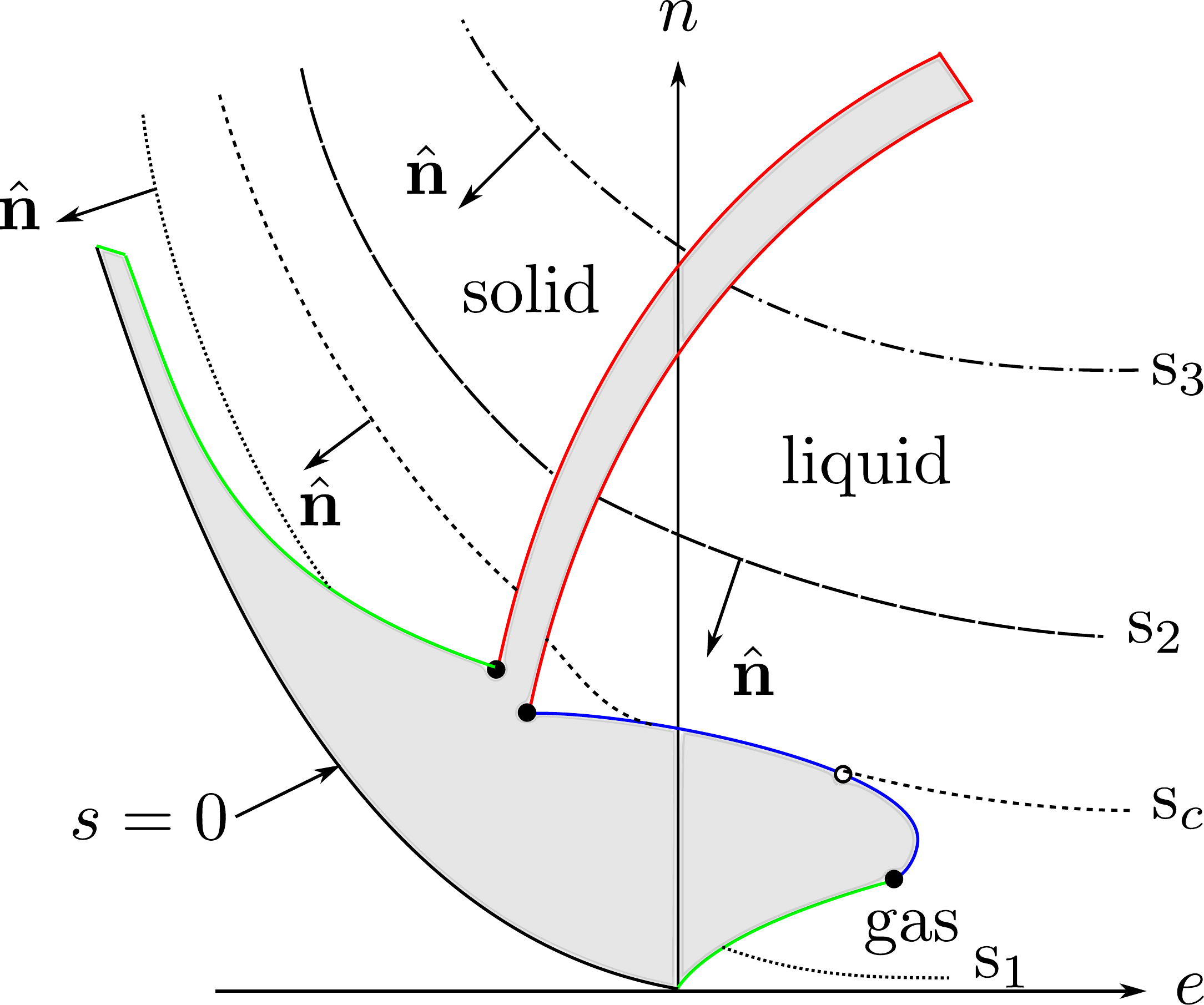}
\caption{A level-curve sketch of $s$ as a function of $e$ and $n$, with $s_1 < s_c < s_2 < s_3$. There are no thermodynamic states in the gray zones. Our interest is in the region near the liquid-gas critical point. Figure taken from Ref. \cite{MML}.}
\label{fig1}
\end{center}
\end{figure}

Let us consider the  entropy  $s = s(e,n)$ of an ``arbitrary'' chemically pure fluid. $s$, $e$ and $n$ are the entropy, energy and  number of particles per unit of volume. By the laws of thermodynamics, $s(e,n)$ is a single valued, concave function of $(e,n)$, such that \cite{LLI,Callen},
\ba
ds & = & \left(\frac{\partial s}{\partial e}\right)_{n} de + \left(\frac{\partial s}{\partial n}\right)_{e} dn  \nonumber \\
& \equiv &  \beta \> de -\alpha \> dn  \label{ds}
\ea
with all the variables in dimensionless units (say, entropy in units of Boltzmann constant and energy and volume with units of two characteristic parameters of intermolecular potentials). $\beta = 1/T$ and $\alpha = \mu/T$, with $T$ the temperature and $\mu$ the chemical potential. By the third law $\beta > 0$, which indicates that for $n =$ constant $s$ is a concave, monotonic increasing function of $e$. Although there is no thermodynamic restriction on $\alpha$, for states near the liquid-vapor transition $\alpha < 0$ \cite{MML} and, as a consequence, $s$ is also a concave, monotonic increasing function of $n$, for $e = $constant, see Fig. \ref{fig1}.\\

The pressure $p$ of the fluid is given by,
\be
\beta p = s + \alpha \>n - \beta \> e\>.
\ee
Thermodynamic equilibrium requires that $\alpha$, $\beta$ and $p$ are {\it continuous} functions of $(e,n)$. In addition, the second law guarantees that the principal curvatures of the surface $s = s(e,n)$ are finite everywhere, except at isolated points such as the critical one. The fact that $s(e,n)$ is concave everywhere yields the stability conditions on the specific heat at constant volume and number of particles $c_v$ and on the isothermal compressibility $\kappa_T$ \cite{LLI,Callen},
\be
-\beta^2 c_v^{-1} =\frac{\partial^2 s}{\partial e^2} < 0\>\label{cn}
\ee
and
\be
- \frac{\beta}{n^2} \kappa_T^{-1} = \frac{\partial^2 s}{\partial n^2} - \frac{\left(\frac{\partial^2 s}{\partial e\partial n}\right)^2}{\frac{\partial^2 s}{\partial e^2}} < 0 \>.\label{kappa}
\ee

\begin{figure}[htbp]
\begin{center}
\includegraphics[width=.45\textwidth]{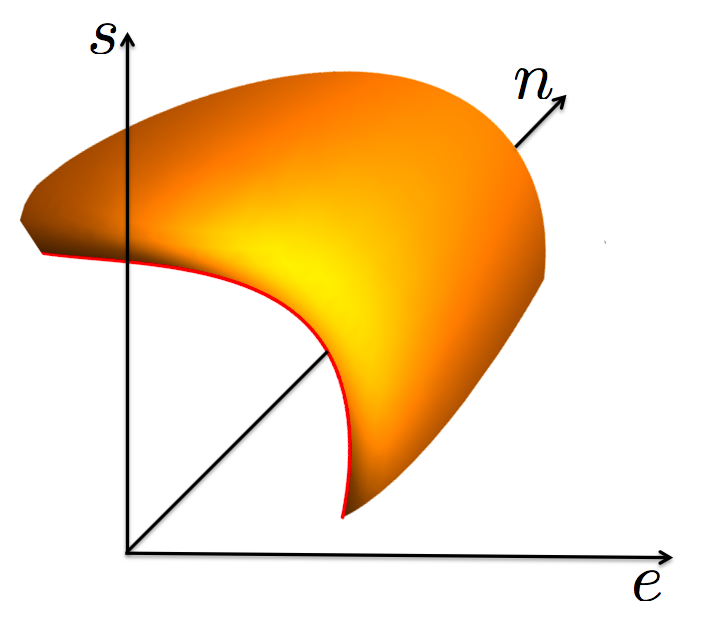}
\caption{A 3D sketch of the function $s = s(e,n)$, in the neigborhood of the critical point, showing the coexistence curve as the edge in red color. }
\label{fig2}
\end{center}
\end{figure}

Now, we consider a fluid that shows a liquid-vapor phase transition ending in a thermodynamic state known as the {\it critical point}, see Fig. \ref{fig1}. In such a phase transition, except at the critical point, there exists a continuum of pairs of  thermodynamic states in equilibrium, with their energy $e$, particle number $n$, and entropy $s$ densities being discontinuous. This physical situation requires that the entropy function $s = s(e,n)$, considered as a surface in a cartesian set of axes $(e,n,s)$, shows a ``cut'', or void, that accounts for the mentioned  discontinuities. The  edge of such a void region is the {\it coexistence} curve, as shown in Fig. \ref{fig2}.  For values of $(e,n)$ ``inside'' the void $s$ is not defined. The curve has a special point, identified as the critical one $(e_c,n_c,s_c)$, such that for a given pair of the mentioned states, one is the liquid phase with values $(e_l,n_l, s_l)$ in one side of the critical point, and the other is the gas phase with $(e_g,n_g,s_g)$ in the opposite side. These states are said to be in coexistence if their temperature $\beta$ and chemical potential $\alpha$ (and so pressure $p$) have the same values. 
As the critical point is approached the two coexisting states coalesce into such a special point. Here, we make the {\it unique} assumption of this discussion, based entirely on experimental and numerical simulations data \cite{Michaels,Heller,Stewart,Mahajan,Panagiotopoulos,Allen,Harris,Binder}: very near the critical point, including the coexistence curve, the surface is symmetric with respect to the plane perpendicular to the tangent at the critical point, as it will be explicitly specified below. This very important empirical observation will lead to the scaling form of $s(e,n)$ and to the well-known critical properties, namely, the necessary divergence of the isothermal compressibility and the possible divergence of the specific heat at constant volumen. A very important consideration is that the entropy surface is an analytic function of $(e,n)$, except at the critical point where it can be non-analytic.
 
\section{An isometric transformation to the critical point}

\begin{figure}[htbp]
\begin{center}
\includegraphics[width=.45\textwidth]{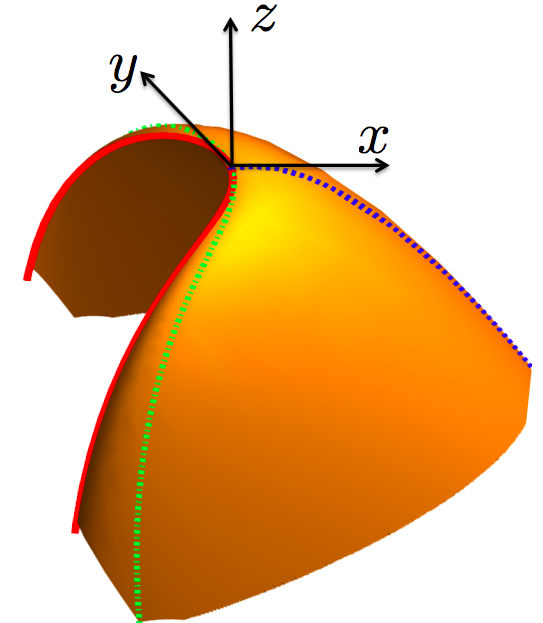}
\caption{(Color online) A sketch of the entropy surface given in Fig. \ref{fig2}, in the local, displaced and rotated axes $\hat m_c$, $\hat t_c$ and $\hat n_c$, defining the axes $x$, $y$ and $z$, respectively. The coexistence curve is in a continuous (red) line, the curve $y = 0$ in a dotted (blue) line and the critical isotherm $\beta_c$ in a dot-dash (green) line. As discussed in Sections IV and VI, the curves $y = 0$ and the coexistence one conform the symmetry-breaking line.}
\label{fig3}
\end{center}
\end{figure}

As mentioned above, the entropy function $s = s(e,n)$ can be considered as a surface $\vec R \equiv (e,n,s(e,n))$ in the right-hand set of axes $(e,n,s)$. Now, for our purposes below, we make a transformation to a cartesian set of axes $(x,y,z)$ located at the  critical $(e_c,n_c,s_c)$, as shown in Fig. \ref{fig3}. The $z-$axis is defined by the normal unit vector at the critical point,
\be
\hat n_c = \frac{(-\beta_c,\alpha_c,1)}{\sqrt{1+\alpha_c^2+\beta_c^2}} \> .\label{nor}
\ee
The $y-$axis in turn is defined by the unit vector tangent to the coexisting curve, $\hat t_c$, and the $x-$axis is then identified by the vector $\hat m_c = \hat t_c \times \hat n_c$, pointing towards the region where the surface $s=s(e,n)$ is defined. 
The relationship between the coordinates $(e,n,s)$ and the local set $(x,y,z)$ is given by,
\ba
x & = & \hat m_c \cdot  \Delta  \vec R  \nonumber \\
y & = & \hat t_c \cdot \Delta  \vec R  \nonumber \\
z & = & \hat n_c \cdot \Delta  \vec R  \>, \label{transf}
\ea
where $\Delta \vec R = (\Delta e, \Delta n, \Delta s)$ with $\Delta e = e-e_c$, $\Delta n = n-n_c$ and $\Delta s = s-s_c$. \\

In the new set of coordinates $(x,y,z)$ the entropy surface can be expressed in terms of the function $z = z(x,y)$, see  third relationship in Eq. (\ref{transf}), and thus is related to the entropy by,
\be
s(e,n) = s_c - \alpha_c \Delta n + \beta_c \Delta e + \sqrt{1 + \alpha_c^2 + \beta_c^2} \> z\left(x,y\right) \label{relat}
\ee
where $y = y(e,n)$ and $x = x(e,n)$ are given by the first two equations of (\ref{transf}), with $\Delta s = \Delta s(e,n)$. \\

With the given transformation, the relations between $\alpha$, $\beta$ and the derivatives of $z$ with respect to $x$ and $y$ are,
\be
\beta = \beta_c + \sqrt{1+\alpha_c^2 + \beta_c^2}\left[ \left(\frac{\partial z}{\partial x}\right)_y \left( \frac{\partial x}{\partial e}\right)_n + \left(\frac{\partial z}{\partial y}\right)_x  \left(\frac{\partial y}{\partial e}\right)_n \right] \>,\label{betaxy}
\ee
\be
-\alpha = -\alpha_c + \sqrt{1+\alpha_c^2 + \beta_c^2}\left[ \left(\frac{\partial z}{\partial x}\right)_y \left( \frac{\partial x}{\partial n}\right)_e + \left(\frac{\partial z}{\partial y}\right)_x  \left(\frac{\partial y}{\partial n}\right)_e \right] \>.\label{alfaxy}
\ee
The following is a general result: from Eq. (\ref{transf}) one finds that the derivatives of $x$ and $y$ with respect to $e$ depend on $\beta$ only, while the derivatives of $x$ and $y$ with respect to $n$, in turn, depend on $\alpha$ only. Therefore, it is a simple exercise to verify that for two coexisting states, the fact that $\alpha$ and $\beta$ have the same value implies that the derivatives $(\partial z/\partial x)_y$ and $(\partial z/\partial y)_x$ must also have the same value for those two coexisting states.\\

Our interest is the description of the entropy in the vicinity of the critical point, namely, for $|x| \ll 1$ and $|y|\ll1$. Following our physical assumption that the coexistence curve is symmetric with respect to the tangent $\hat t_c$ at the critical point, we can assert that in this set of coordinates the coexistence curve is given by a relationship between $x$ and $y$, namely $x = x_{\rm coex}(y)$, such that the coexistence curve can be written parametrically through the vector,
\be
\vec R_{\rm coex}(y) \approx \left(x_{\rm coex}(y),y, z(x_{\rm coex}(y),y)\right)  \>\>\>\>\textrm{for}\>\>\>|y| \ll 1 \>,\label{Rcoex}
\ee
with $x_{\rm coex}(y)$ a symmetric function of $y$,
\be
x_{\rm coex} \approx - c_0 \left(y^{2}\right)^\Delta \>\>\>\>\textrm{for}\>\>\>|y| \ll 1 \>,\label{coex}
\ee
with $c_0 >0$, a coefficient characteristic of the given fluid. It is physically reasonable to assume that $\Delta \ge 1$, for the coexistence curve to have a curvature either finite or zero at the critical point. Since $\Delta$ is not limited to be an integer, the coexistence curve can be non-analytic at the critical point. The coexistence curve is represented by the solid (red) line in Fig. \ref{fig3}. From now on, we shall use the notation $y^{2 \eta} \equiv (y^2)^\eta$ for any $\eta$, to avoid cumbersome expressions.\\ 

The assumed form of the coexistence curve, Eq.(\ref{coex}), implies that in the vicinity of the critical point the thermodynamic states $(x_{\rm coex}(y), y)$ and $(x_{\rm coex}(y), -y)$ coexist. Hence, the temperature $\beta$ and the  chemical potential $\alpha$ have the same value at those states, that is, $\beta(x_{\rm coex}(y), y) \approx \beta(x_{\rm coex}(y), -y)$ and  $\alpha(x_{\rm coex}(y), y) \approx \alpha(x_{\rm coex}(y), -y)$. 
Now we make the most important assumption of the present work: very near the critical point, that is, at a leading order, the function $z = z(x,y)$ is symmetric on $y$, $z(x,y) \approx z(x,-y)$. As mentioned in the Introduction, one can show that the van der Waals model indeed satisfies this requirement. From this assumption follows a trascendental result. First note that 
the even symmetry of $z$ on $y$ implies that the derivative of $z$ with respect to $y$ is odd. However, as stated above, both derivatives of $z$ with respect to $x$ and $y$ must be equal at coexistence states. Therefore, since the derivative of $z$ with respect to $y$ at coexistence must be both odd and even, this can only be true if it vanishes at all points in the coexistence curve, that is, 
\be
\left. \left(\frac{\partial z}{\partial y}\right)_x \right|_{\rm coexistence} = 0 \>\>\>\>\textrm{for}\>\>\>|y| \ll 1 \>.\label{condicion}
\ee
As we now show, this condition on the shape of the surface is so strong that it implies that $s(e,n)$ must obey scaling.\\

\section{The scaling form of the entropy}  

A very important condition is that the entropy surface is analytic everywhere, except perhaps at the critical point. Therefore, we can make an $x$-power expansion of $z=z(x,y)$ around $x = 0$, for an arbitrary value of $y \ne 0$, near the critical point. This yields,
\be
z(x,y) \approx -\sum_{n=0}^\infty f_n(y^2) x^n\label{eq_series_approximation}
\ee
where the functions $f_n(y^2)$ need not be analytic at $y = 0$. Let us now take the derivative of $z$ with respect to $y$,
\be
 \frac{\partial z}{\partial y} = -2 y \sum_{n=0}^\infty f_n^\prime (y^2) x^n \label{proposal}
 \ee
 where the prime means differentiation with respect to the argument. As we have just argued above, see Eq. (\ref{condicion}), this derivative must be zero at the coexistence curve, that is, for $x \approx -c_0 \> y^{2\Delta}$,
\be
y \sum_{n=0}^\infty f_n^\prime (y^2) \left(-c_0 \> y^{2\Delta}\right)^n = 0 \>\>\>\>\forall \> y \ne 0 \>.\label{req}
\ee
We note that this condition imposes a very strong restriction on the functions $f_n(y^2)$ since the equality must be true 
for a continuum of values of $y \ne 0$. Although we are considering this to hold near the critical point only, we can write a quite general expression that satisfies the above requirement. That is, a general solution to Eq.(\ref{req}) is that $f_n(y^2)$ is a power law expansion, not necessarily analytic:
\be
f_n(y^2) = A_n y^{2\Gamma_n} + B_n y^{2\Xi_n} + C_n y^{2\Omega_n} + \cdots \label{req2}
\ee
where the exponents $\Gamma_n$, $\Xi_n$, $\Omega_n$, and so on, are not integers in general. With this proposal we find that for expression (\ref{req}) to hold, these exponents must satisfy,
\ba
\Gamma_n + n \Delta &=& \Gamma_0 \nonumber \\
\Xi_n +  n \Delta &=& \Xi_0 \nonumber \\
\Omega_n +  n \Delta &=& \Omega_0 \label{req3}
\ea
and so on, with $\Gamma_0 < \Xi_0 < \Omega_0$, etcetera, with no loss of generality. To verify it, substitute the above into (\ref{req}),
\ba
\frac{y^{2\Gamma_0}}{y} \sum_{n=0}^\infty \Gamma_n A_n (-c_0)^n + \frac{y^{2\Xi_0}}{y} \sum_{n=0}^\infty \Xi_n B_n (-c_0)^n + \frac{y^{2\Omega_0}}{y} \sum_{n=0}^\infty \Omega_n C_n (-c_0)^n + \dots = 0 \>,
\ea
from which one concludes that to satisfy the equality for all values of $y$, each sum  must vanish separately. In particular, and our interest here, it must be true that,
\begin{equation}
\sum_{n=0}^\infty \Gamma_n A_n (-c_0)^n = 0 \>.
\end{equation}
Hence, substituting Eqs. (\ref{req2}) and (\ref{req3}) into (\ref{eq_series_approximation}), we can write,
\begin{equation}
z(x,y) \approx - y^{2\Gamma_0} \sum_{n=0}^\infty A_n \left(\frac{x}{y^{2\Delta}}\right)^n - y^{2\Xi_0}\sum_{n=0}^\infty B_n \left(\frac{x}{y^{2\Delta}}\right)^n - y^{2\Omega_0}\sum_{n=0}^\infty C_n \left(\frac{x}{y^{2\Delta}}\right)^n - \dots\>\>\>\>\forall \> y \ne 0 \>.
\end{equation}

The above form shows a general functional dependence that obeys scaling. However, since we have limited ourselves to assumptions and approximations very near the critical point, we keep the lowest order of $f_n(y^2)$ only. This yields the desired scaling expression for the surface function $z(x,y)$ in terms of two unknown exponentes $\Delta$ and $\Gamma_0$,
\ba
z(x,y) &\approx&  - y^{2\Gamma_0} \sum_{n=0}^\infty A_n \left(\frac{x}{y^{2\Delta}}\right)^n  \nonumber \\
&\equiv&  - y^{2\Gamma_0} {\cal F}\left(\frac{x}{y^{2\Delta}}\right) \>\>\>\>\>\textrm{for}\> y \ne 0 \>,\label{scal}
\ea
where in the last line we have defined the scaling function ${\cal F}(X)$, with $X = x/y^{2\Delta}$, which by construction is an analytic function of its argument. This demonstrates that the entropy can be written in a scaling form in terms of the variables $x$ and $y$ \cite{Widom1965,Ma,review-SH}. In the following section we analyze the predictions of this finding and in Appendix A we show that the series expansion given in Eq. (\ref{eq_series_approximation}) can be made even more general, leading too to the scaling form in Eq. (\ref{scal}). But before writing the full expression of the entropy, we first address how to deal with values of $y = 0$ and $x \ne 0$; this is discussed in many reviews, such as Ref. \cite{review-SH}.\\

Because $z(x,y)$ is a maximum at the origin, then ${\cal F}(x/y^{2\Delta}) > 0$. For $x = 0$, $z \approx - A_0 y^{2 \Gamma_0}$, where ${\cal F}(0) = A_0 >0$. While the scaling function is valid everywhere {\it on} the entropy surface, it can be explicitly evaluated for $x \gtrsim -c \>y^{2\Delta}$ but $y \ne 0$ only, as expressed above. However, for $x > 0$ and $y = 0$, $z$ must be a function of $x$ only. Therefore, asymptotically, the scaling function must behave as,
\be
{\cal F} \left(\frac{x}{y^{2\Delta}} \right) \approx B_0 \left(\frac{x}{y^{2\Delta}} \right)^{\Gamma_0/\Delta} \>\>\>\>\textrm{for}\>\>\frac{x}{y^{2\Delta}} \to \infty \iff x \ll1 \>\>\>{\rm and}\>\>\> y \to 0 \>,
\ee
with $B_0 > 0$, such that, see (\ref{scal}), $z(x,0) \approx - B_0 x^{\Gamma_0/\Delta_0}$, a function of $x$ only. This indicates that, in general \cite{review-SH}, as long as $x >0$, we can  ``invert'' the function ${\cal F}$ by identifying a new function as,
\be
{\cal G}\left(\frac{y^2}{x^{1/\Delta}}\right) \equiv \left(\frac{y^2}{x^{1/\Delta}}\right)^{\Gamma_0} {\cal F}\left(\frac{x}{y^{2\Delta}} \right) \>\>\>\>\textrm{for}\>\>x \ge 0 \>,
\ee
and since $z(x,y)$ can be non analytic at criticality, this scaling function should also be an infinite series,
\be
{\cal G}\left(\frac{y^2}{x^{1/\Delta}}\right) = \sum_{n=0}^\infty B_n \left(\frac{y^2}{x^{1/\Delta}}\right)^n \>.
\ee
Conversely and for consistency, asymptotically it must be true that, 
\be
{\cal G}\left(\frac{y^2}{x^{1/\Delta}}\right)   \approx  A_0 \left(\frac{y^2}{x^{1/\Delta}}\right) ^{\Gamma_0} \>\>\>\>\textrm{for}\>\>\frac{y^2}{x^{1/\Delta}}  \to \infty \iff |y| \ll1 \>\>\>{\rm and}\>\>\> x \to 0^+ \>. \>
\ee
Therefore, one can write $z(x,y)$ in terms of this complementary scaling function, 
\be
z(x,y) \approx - x^{\Gamma_0/\Delta} \> {\cal G}\left(\frac{y^2}{x^{1/\Delta}}\right)  \>\>\>\>\textrm{for}\>\>x > 0. \label{G2}
\ee
This form is particularly useful to calculate properties of the surface in the  limit $x \to 0^+$ for $y = 0$, namely, at the states on the dotted (blue) line in Fig. \ref{fig3}, that  we can now identify as  the line of ``symmetry breaking'' \cite{Ma,Amit}. This is because we can now show that the derivative of $z$ with respect to $y$, at $y = 0$ and for any $x > 0$, vanishes:
\begin{equation}
\left. \frac{\partial z}{\partial y}\right|_{y = 0} = - \left. 2 y \> x^{(\Gamma_0-1)/\Delta } \> {\cal G}^\prime\left(\frac{y^2}{x^{1/\Delta}} \right)\right|_{y=0} = 0 \>.\label{crit-line}
\end{equation}
 That is, for $x > 0$ and $y = 0$ there is only one phase (a so-called supercritical fluid) with the derivative $(\partial z/\partial y)$ being zero; then for $x < 0$ but at the coexistence curve, there are two phases with $(\partial z/\partial y)$ remaining zero. The analogy is with the para-ferromagnetic case where the line of symmetry breaking is when the magnetic field vanishes \cite{Ma,yo}; that is, in the local frame $(x,y,z)$ the derivative of $z$ with respect to $y$ is the analog of the magnetic field. We return to this relation in the final section.\\

 With the previous identification of the scaling function we can write explicit forms of the entropy in the vicinity of the critical point,
 \be
s(e,n) \approx s_c - \alpha_c \Delta n + \beta_c \Delta e - \sqrt{1 + \alpha_c^2 + \beta_c^2} \> y^{2\Gamma_0} {\cal F}\left(\frac{x}{y^{2\Delta}}\right) \label{sf}
\ee
valid for all values of $x \gtrsim - c_0 \> y^{2\Delta}$ and $y \ne 0$, or its alternative form
\be
s(e,n) \approx s_c - \alpha_c \Delta n + \beta_c \Delta e - \sqrt{1 + \alpha_c^2 + \beta_c^2} \> x^{\Gamma_0/\Delta} {\cal G}\left(\frac{y^2}{x^{1/\Delta}}\right)  \>,\label{sg}
\ee
valid for $x > 0$ and all values of $y$. Consistently, $x$ and $y$, given by (\ref{transf}) in terms of $\Delta e$ and $\Delta n$ should also be expanded up to the appropriate order to yield the leading order of the thermodynamic properties in question.\\

\begin{figure}[htbp]
\begin{center}
\includegraphics[width=.6\textwidth]{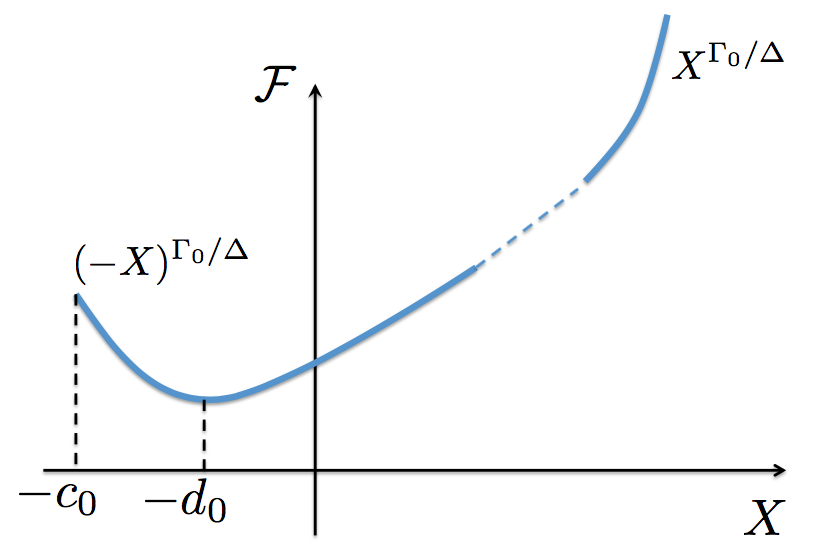}
\caption{(Color online) Sketch of the entropy scaling function ${\cal F}$ as a function of its argument $X$.  This function is assumed to be analytic at $X = 0$ and defined for $-c_0 \le X < \infty$, with $X =-c_0$ occurring at the coexistence curve. It has the asymptotic limit ${\cal F} \sim X^{\Gamma_0/\Delta}$ for $X \gg 1$. The function is a minimum at $X = -d_0$ which corresponds to the critical isotherm. In section VI we discuss the asymptotic value ${\cal F} \sim (-X)^{\Gamma_0/\Delta}$ as $X \to -c_0$.}
\label{sketch}
\end{center}
\end{figure}

To conclude this section, we can observe general properties of the scaling functions. Since $z(x,y)$ is a maximum at $(x,y) = (0,0)$, then ${\cal F} > 0$ and ${\cal G} > 0$ for all values of their arguments. For positive values of $x/y^{2\Delta}$, ${\cal F}^\prime>0$. However, at coexistence  ${\cal F}^\prime(-c_0)$ must be negative; as it will be seen below, this is due to the fact that the (inverse) temperature $\beta$ at coexistence states is greater than the critical (inverse) temperature $\beta_c$. As a consequence, at a certain negative value of its argument ${\cal F}^\prime(-d_0)=0$ where it changes sign. As we will show below, ${\cal F}^\prime=0$ occurs at the critical isotherm. See Fig. \ref{sketch} for a sketch of  ${\cal F}$.

\section{Critical properties and exponents} 

Empirical evidence shows that the thermodynamic behavior of any fluid near the critical point is characterized by power-law dependences on temperature and density, with well defined critical exponents. As we now revise, the scaling form obtained above yields those relationships. In particular, there are four expressions that condense such a critical behavior. Since the analysis leading to the identification of the critical exponents is quite lengthy, we quote here the main results and leave the details for Appendix B.\\

{\bf Order parameter.} The first relationship is the evaluation of the coexistence curve in terms of the temperature and the particle density. This is the so-called equation for the order parameter below the critical temperature \cite{Ma,review-SH}. For this, we use Eq. (\ref{betaxy}) for the temperature $\beta$, the conditions at coexistence that $x \approx - c_0 y^{2\Delta}$ and that the derivative of $z$ with respect to $y$ vanishes. Evaluation to lowest significant order leads to
\begin{equation}
\beta - \beta_c \approx - \sqrt{1+\alpha_c^2+\beta_c^2} {\cal F}^\prime(-c_0) \hat m_c \cdot \vec \tau_{\beta_c} \left(\hat t_c \cdot \left[ \vec \tau_{\alpha_c} -   \vec \tau_{\beta_c} \frac{\hat m_c \cdot \vec \tau_{\alpha_c}}{\hat m_c \cdot \vec \tau_{\beta_c}} \right] \Delta n\right)^{2\Gamma_0-2\Delta} \>,\label{betadn}
\end{equation}
where one verifies that $ {\cal F}^\prime(-c_0) < 0$. Here we have introduced two vectors,
\begin{eqnarray}
\vec \tau_{\beta_c} &=& (1,0,\beta_c) \>\nonumber \\
\vec \tau_{\alpha_c} &=& (0,1,-\alpha_c) \>,
\end{eqnarray}
that, incidentally, are tangent to the entropy surface $s(e,n)$ at the critical point, as $\hat t_c$ and $\hat m_c$ also are, see Appendix B. Since to leading order $\beta - \beta_c \approx (T_c - T)/T_c^2$, from Eq. (\ref{betadn}) one can write $T_c - T \approx  (\Delta n/{\cal A})^{2\Gamma_0-2\Delta}$, where the constant ${\cal A}$ can be read off the equation. Hence, inverting, one finally obtains,
\begin{equation}
|\Delta n| \approx {\cal A} \left(T_c - T\right)^{1/(2\Gamma_0-2\Delta)} \>.\label{dnbetaexp}
\end{equation}
This is the equation of the coexistence curve in terms of temperature and density. From this expression one identifies the critical exponent ``beta'' \cite{Fisher-review,Ma,review-SH},
\begin{equation}
\hat \beta = 1/(2\Gamma_0-2\Delta) \>.\label{betaexp}
\end{equation}
We have denoted this exponent with a ``hat'' in order to avoid confusion with the inverse of the temperature. We will use the same notation for the other critical exponents. \\

{\bf Critical isotherm.} We now turn our attention to the relationship between the chemical potential $\alpha$ and the particle density $n$ along the critical isotherm $\beta = \beta_c$, near the critical point. As it is detailed in Appendix B, this requires a very careful analysis. The main point is the identification of the relationship between $x$ and $y$ along the critical isotherm. For this we quote the general condition obeyed by $x$ and $y$ by setting $\beta = \beta_c$ in Eq. (\ref{betaxy}):
\begin{equation}
0 =  \vec \tau_{\beta_c} \cdot \left[\hat m_c \left(\frac{\partial z}{\partial x} \right)_y+\hat t_c \left(\frac{\partial z}{\partial x} \right)_y\right]  \Bigg|_{\beta_c}  \>.\label{condiso}
\end{equation}
This expression is exact and the derivatives are assumed to be evaluated at the critical isotherm. Then, using the scaling form for the function $z(x,y)$, Eq. (\ref{scal}), a systematic expansion in (non-integer) powers of $y$ in Eq. (\ref{condiso}) should be found, see Appendix B. In such an expansion each coefficient in the series must vanish. For this to be true, the critical isotherm must be of the form $x \approx - d_0 y^{2\Delta}$, defined by the condition ${\cal F}^\prime(-d_0) = 0$, see Fig. \ref{sketch}, with $d_0$ a constant coefficient that must obey $d_0 < c_0$ in order to be within the surface, as also indicated by the dotted-dash (green) line in Fig. \ref{fig3}. That is, the critical isotherm has the same mathematical form of the coexistence curve, except that with a smaller coefficient. To the best of our knowledge this result has not been pointed out before. Now, using these results into equation (\ref{alfaxy}) for $\alpha$, see Appendix B, yields an expression in terms of $y$,
\begin{equation}
\alpha \approx \alpha_c - 2 \Gamma_0 \sqrt{1+\alpha_c^2 + \beta_c^2} \> \vec \tau_{\alpha_c} \cdot \left[\hat t_c -  \frac{\vec \tau_{\beta_c}\cdot \hat t_c}{\vec \tau_{\beta_c}\cdot\hat m_c}\hat m_c\right] {\cal F}(-d_0) \> y^{2\Gamma_0 -1} \>.\label{alfay}
\end{equation}
\medskip 
With the use of Eqs. (\ref{transf}) and the critical isotherm relationship $x \approx - d_0 y^{2\Delta}$, to leading order of approximation we can express $x$ and $y$ in terms of $\Delta e$ and $\Delta n$, allowing to solve for $y$ in terms of $\Delta n$. The final result is,
\begin{equation}
\alpha \approx \alpha_c - 2 \Gamma_0 \sqrt{1+\alpha_c^2 + \beta_c^2} \> \vec \tau_{\alpha_c} \cdot \left[\hat t_c -  \frac{\vec \tau_{\beta_c}\cdot \hat t_c}{\vec \tau_{\beta_c}\cdot\hat m_c}\hat m_c\right] {\cal F}(-d_0) \> \left(\hat t_c \cdot \left[ \vec \tau_{\alpha_c} -   \vec \tau_{\beta_c} \frac{\hat m_c \cdot \vec \tau_{\alpha_c}}{\hat m_c \cdot \vec \tau_{\beta_c}} \right] \Delta n\right)^{2\Gamma_0-1} \>.
\end{equation}
This expression indicates that for $\beta = \beta_c$, as $\alpha \to \alpha_c$, $n \to n_c$ with an exponent that we identify as the ``delta exponent'' $\tilde \delta$ \cite{Fisher-review,Ma,review-SH},
\begin{equation}
\hat \delta = 2 \Gamma_0 - 1 \>.\label{deltaexp}
\end{equation}

{\bf Specific heat at constant volume.} As it is known, the specific heat $c_v$ and the isothermal compressibility $\kappa_T$  show a characteristic behavior as the critical point is approached. Although there are an infinite numbers of paths to approach it, we will use a ``usual'' one, which is the curve $ \Delta n = 0$ and let $T \to T_c$ from above. This corresponds to states with $x >0$ only and, therefore,  we use the scaling form of $z$ given by Eq. (\ref{G2}) in terms of the ${\cal G}$ scaling function. As derived in Appendix B, very near the critical point the condition $\Delta n = 0$ implies a linear relationship between $x$ and $y$,
\be
y \approx \frac{\hat t_c \cdot \vec \tau_{\beta_c}}{\hat m_c \cdot \vec \tau_{\beta_c}} x \>.
\ee
This allows us to make an expansion of both $c_v$ and $\beta$ in terms of $x$ solely, see Eqs. (\ref{cn}) and (\ref{betaxy}), such that one can solve for the later and find a relationship between the former. One gets, see Appendix B,
\be
\beta^2 c_v^{-1} \approx \left(\frac{\Gamma_0}{\Delta}-1\right) \left(\sqrt{1+\alpha_c^2 + \beta_c^2} \>\hat m_c \cdot \vec \tau_{\beta_c} \frac{\Gamma_0}{\Delta} {\cal G}(0) \right)^{\frac{\Delta}{\Gamma_0-\Delta}}\left(\beta_c -\beta\right)^{\frac{\Gamma_0-2\Delta}{\Gamma_0-\Delta}} \>.\label{cnfinal}
\ee
With this expression we can read the critical exponent ``alpha'' \cite{Fisher-review,Ma,review-SH}
\be
\tilde \alpha = \frac{\Gamma_0-2\Delta}{\Gamma_0-\Delta} \>.\label{alfaexp}
\ee
Expression (\ref{cnfinal}) indicates that it must be true that $\Gamma_0 \ge 2 \Delta$, otherwise the specific heat would become zero at the critical point, an unphysical result in violation of the second law of thermodynamics. This condition ensures that the entropy function $s = s(e,n)$ is smooth everywhere. Since $\Delta \ge 1$, we conclude that $\Gamma_0 \ge 2$, an important result to be used below.\\

{\bf Isothermal compressibility.} As mentioned above, there are an infinity ways to approach the critical point on the surface $s(e,n)$. Hence, for the elucidation of the behavior of the isothermal compressibility $\kappa_T$ we choose again the curve $\Delta n = 0$ and let $T \to T_c$ from above. This is the lengthiest of the critical exponents calculations and the main steps are given in Appendix B. Once more, we exploit the linear relationship between $y$ and $x$ to obtain
\be
\frac{\beta}{n^2}\kappa_T^{-1}  \approx  {\cal B} \> \left(1+\alpha_c^2 + \beta_c^2\right)^{\frac{1-\Delta}{2(\Gamma_0-\Delta)}}(\beta_c - \beta)^{\frac{\Gamma_0 -1}{\Gamma_0-\Delta}}\>,\label{kappafinal}
\ee
where the coefficient ${\cal B}$ is given by,
\be
{\cal B}  = 2\>{\cal G}^\prime(0) \left(\hat t \cdot \vec \tau_{\alpha_c} \right)^2 \left[ 1 -  \frac{\left(\hat t_c \cdot \vec \tau_{\beta_c} \right)}{ \left(\hat t_c \cdot \vec \tau_{\alpha_c} \right) }\frac{\hat m_c \cdot \vec \tau_{\alpha_c}}{\hat m_c \cdot \vec \tau_{\beta_c}}\right]^2\>\left[\frac{\Delta}{\Gamma_0 \hat m_c \cdot \vec \tau_{\beta_c}  {\cal G}(0)}\right]^{\frac{\Gamma_0-1}{\Gamma_0-\Delta}}\>.
\ee
From Eq. (\ref{kappafinal}) we read the critical exponent ``gamma'' \cite{Fisher-review,Ma,review-SH},
\be
\hat \gamma = \frac{\Gamma_0 -1}{\Gamma_0-\Delta} \>,\label{gammaexp}
\ee
indicating that, since $\Gamma_0 \ge 2$, the isothermal compressibility {\it must} diverge at the critical point. We discuss this result further below.\\

By reading the critical exponents $\hat \alpha$, $\hat \beta$, $\hat \gamma$ and $\hat \delta$ from this section, one sees that all depend on the so-far unknown exponents $\Delta \ge 1$ and $\Gamma_0 \ge 2$. It is then a simple exercise to verify that they obey the so-called Rushbrooke \cite{Rushbrooke} and Griffiths \cite{Griffiths} equalities, $\hat \alpha + 2 \hat \beta + \hat \gamma =2$ and $\hat \beta (1 + \hat \delta) = 2 - \hat \alpha$, originally predicted as inequalities and shown to be equalities by Widom with the scaling hypothesis \cite{Widom1965}.

\section{Final remarks} 

The derivation of the scaling form of the entropy here presented partialy follows the study of Ref. \cite{yo} for a ferromagnetic system. However, due to the much more complicated structure of the entropy function of a fluid near its critical point, the study here includes the magnetic one as an special case. In that situation, the entropy per unit volume $s$ is a function of the energy per unit volume $e$ and the magnetization per unit volume $m$, that is, $s = s(e,m)$. By the laws of thermodynamics $s$ is a concave function of  those variables. In this case $ds =(1/T) de +  (H/T) dm$ where $H$ is the magnetic field. By physical reasons of symmetry, $s$ is an even function of $m$ and, therefore, $H = H(e,m)$ is an odd function of $m$. The critical point is at $\beta=\beta_c$, $m = 0$ and $H = 0$. Thus, the symmetry breaking occurs for $T < T_c$ where, with $H = 0$, one finds $m \ne 0$. Hence, since $s$ is even in $m$ and the critical value of $H$ is zero, the identification $y = m$ in our discussion follows right away. This suggests that the isometric transformation (\ref{transf}) is actually unnecessary, yielding $x = \Delta e$. Therefore, one immediately finds that the entropy can be written near the critical point as,
\be
s(e,m) \approx s_c+\frac{1}{T_c} \Delta e - m^{2\Gamma_0} {\cal F}\left(\frac{\Delta e}{m^{2\Delta}}\right) \>,
\ee
with $\Delta e \approx - c_0 \> m^{2 \Delta}$ the coexistence curve and ${\cal F}$ a scaling function \cite{yo}. It is worthwhile to recall that when analogies between magnetic and fluid systems are considered, see e.g. Refs. \cite{Fisher-review} and \cite{Ma}, the magnetization $m$ is the analog of density $\Delta n$ and it appears natural to identify the magnetic field $H$ as the analog of the chemical potential $\Delta \mu$ (or the pressure) since these are the thermodynamic conjugate variables of the former. However, the present study shows that this is not the case. That is, the correct analogy of the  magnetic symmetry-breaking states with $H = 0$, are in a fluid those for which the derivative $(\partial z/\partial y)$ vanishes, given by Eq. (\ref{crit-line}), as discussed in Section II. That is, very near but above the critical point, Eq. (\ref{crit-line}) leads to the symmetry-breaking line in the fluid, defined by
\begin{equation}
\alpha - \alpha_c \approx - (\beta - \beta_c) \frac{\hat m_c \cdot \vec \tau_{\alpha_c}}{\hat m_c \cdot \vec \tau_{\beta_c}} \>,
\end{equation}
while below the critical point, the line is the coexistence curve $x \approx - c_0 \> y^{2\Delta}$, or the expression given by Eq. (\ref{dnbetaexp}) in terms of density and temperature. As a matter of fact, this was already implicitly discussed by Widom in Ref. \cite{Widom1965}.\\

Although not shown here, an illustrative and pedagogical exercise  is the analysis of the mean-field van der Waals fluid \cite{LLI}, in the light of the present study. In this case one can find the explicit isometric transformation given by Eq. (\ref{transf}) and work out and verify all the predictions here discussed. As one can expect, van der Waals and ferromagnetism Landau mean field models correspond to $\Gamma_0 =2$ and $\Delta = 1$. Details will be given elsewhere.\\

We find highly interesting to point out that while the divergence of the compressibility at the critical point is a physical result with profound and trascendental consequences, notably the foundation of the renormalization group description of the critical point, it appears here to follow as a geometric constraint of the entropy surface at the critical point. That is, the condition $\Gamma_0 \ge 2$, which ultimately follows from the second law of thermodynamics, indicates ``simply'' that the gaussian curvature of the entropy surface, being positive and finite at any stable thermodynamic state, becomes zero at the critical point. This implies that, at least, one of the eigenvalues of the surface curvature tensor vanishes at the critical point, if not both. 
But this general result can be obtained from pure geometrical arguments without appealing to scaling. In other words, both scaling and the flatness of the entropy surface at the critical point, follow from the equilibrium conditions at coexistence and from the fact that there is an ending point to such a coexistence.  To provide evidence for this, 
first we recall that the equilibrium conditions of having the same temperature and chemical potential (and thus pressure) for a given pair of coexistence states, is equivalent to assert that while 
the entropy is discontinuous at the first order phase transition coexistence curve, the surface normal vector $\hat n = (1,-\beta,\alpha)/\sqrt{1+\alpha^2+\beta^2}$ is the same at both coexisting states. Therefore, since such a pair of normal vectors must coalesce to the critical normal $\hat n_c$ staying parallel between them, it is not difficult to show that this implies that the surface necessarily becomes flat at the critical point, 
at least along one of its principal directions.  The case where only one of the curvature eigenvalues is zero corresponds to $\Gamma_0 = 2$ and $\Delta = 1$, the mentioned mean-field case to which Landau and van der Waals theories belong \cite{Ma,review-SH}. For $\Gamma_0 > 2\Delta$ both eigenvalues are explicitly zero. There is a borderline case, not considered here but amply discussed in the seminal paper by Widom \cite{Widom1965}, in which $\Gamma_0 = 2\Delta \ne 2$, such as the two-dimensional Ising model \cite{Fisher-review,Onsager}, where $c_v \to \infty$ logarithmically at the critical point, and both eigenvalues also vanish. One can further verify that the critical exponents for this case correspond to $\Gamma_0 = 8$ and $\Delta = 4$.\\

To conclude we would like to speculate about going further with a pure thermodynamics search for the elucidation of the actual values of the critical exponents. As we have discussed here, from the laws of thermodynamics and the reasonable assumption of a local even symmetry in $y$ in the vicinity of the critical point, scaling follows as a consequence of the phase coexistence requirement. But the most that we can conclude so far is that there exist only two seemingly  independent ``unknown'' exponents $\Delta \ge 1$ and $\Gamma_0 \ge 2$. However, as elaborated above, scaling has also a root in the geometric constraint that a coexistence curve imposes on the necessary discontinuity of the entropy surface. Hence, a question arises as to whether there could be additional constraints that would force the exponents $\Delta$ and $\Gamma_0$ not to be independent. The truthfulness of this would imply a very deep consequence, for then only one exponent would be necessary to find out, the rest of them following by scaling. And we insist, this by pure thermodynamics without appealing to an atomic statistical physics description. Hence, we highlight here once more that the coexistence condition that leads to scaling is the requirement that the derivative of the function $z$ with respect to $y$ vanishes at the coexistence curve, $x \approx - c_0 y^{2\Delta}$. This yields the interesting condition, see Eq. (\ref{scal}),
\be
\frac{\Gamma_0}{\Delta} = - c_0 \left. \frac{d}{d X} \ln  {\cal F}(X)\right|_{-c_0} \> .\label{condF}
\ee
This is an unexpected result in the context of the universality of the exponents $\Gamma_0$ and $\Delta$, since it is usually believed that their values should be independent of the precise functional form of the scaling function ${\cal F}(X)$. The above expression apparently indicates that $\Gamma_0$ and $\Delta$ are not independent, although it may just be an identity showing the asymptotic behavior of ${\cal F}$ near the coexistence curve, as illustrated in Fig. \ref{sketch}. We believe it is worthwhile to explore possible forms of the scaling function in terms of concave surfaces with discontinuities, to find out whether it is indeed an identity or it is a gate to find an additional thermodynamic critical exponents relationship.

\acknowledgments{This work was partially funded by grant PAPIIT-UNAM IN108620.  JCO thanks CONACYT for a graduate studies scholarship and DO for a SNI-III assistant scholarship.}

\section*{Appendix}

\subsection{A generalization of the scaling form of $z$, Eq. \eqref{eq_series_approximation}}

A generalization of the series approximation of the function $z(x,y)$ can be made by replacing the $n$ index in equation \eqref{eq_series_approximation} for a strictly increasing function $\omega(n)$. This could allow for non-analytic scaling functions, although with a restriction that would also permit the entropy to have finite curvatures at all points, except the critical point. An example of this kind of functions is $\omega(n) = \gamma + \theta n$, where $\gamma$ and $\theta$ are constants. Then in the same way as presented above, we can write
\begin{equation}
z(x,y)= - \sum_{n=0}^{\infty} f_n(y^2)x^{\omega (n)}.
\end{equation}
Also we can express each function $f_n(q)$ as
\begin{equation}
f_n(q) = \sum_{i=1}^{\infty}A_{ni} q^{\Gamma_{ni}},
\end{equation}
where $\Gamma_{ni} < \Gamma_{n (i+1)}$. At the coexistence curve $ x \approx - c_0 y^{2\Delta}$ we have $(\partial z / \partial y)_{x}=0$, so assuming a well behavior of this double sum we find
\begin{equation}\label{eq_key_step}
0 = \frac{1}{y} \sum_{i=1}^{\infty} \sum_{n=0}^{\infty} 2 A_{ni}\Gamma_{ni} (-c_0)^{\omega (n)} y^{2(\Gamma_{ni}+\Delta \omega(n) )},
\end{equation}
where we have interchanged the order of summation. 
From equation \eqref{eq_key_step} we find a relation between the exponents given by
\begin{equation}\label{eq_exponents_index_restriction}
\Gamma_{0i} = \Gamma_{ni} + \Delta \omega(n),
\end{equation}
and a restriction over the coefficients,
\begin{equation}\label{eq_coefficients_restriction}
\sum_{n=0}^{\infty} A_{ni}\Gamma_{ni} (-c_0)^{\omega (n)}=0.
\end{equation}
Substituting the relation \eqref{eq_exponents_index_restriction} in $z(x,y)$ gives, with the lowest order $\Gamma_{n0}$ exponent,
\begin{equation}
z(x,y) \approx - (y^2)^{\Gamma_{01}} \sum_{n=0}^{\infty} A_{n1}  \Bigg( \frac{x}{y^{2\Delta}}  \Bigg)^{\omega (n)} + \cdots\>,
\end{equation}
where the scaling form can be identified. 

\subsection{Calculation of critical exponents}

In this Appendix we provide some of the important details to calculate the critical exponents presented in section V. First, we recall that the basic expressions are those for the temperature $\beta$ and chemical potential $\alpha$ given in Eqs. (\ref{betaxy}) and (\ref{alfaxy}). These provide relationships between the previos variables and the derivatives of the function $z(x,y)$ with respect to its variables. These must be used in conjunction with the transformation given by Eqs. (\ref{transf}) and, certainly, the scaling forms of $z(x,y)$ given by Eqs. (\ref{scal}) and (\ref{G2}).\\

It is convenient to define the vectors $\vec \tau_\beta = (1,0,\beta)$ and $\vec \tau_\alpha = (0,1,-\alpha)$, tangents to a given point $(e,n)$ 
on the surface (which incidentally define the local metric and curvature tensors). With these, one finds,
\ba
\left( \frac{\partial x}{\partial e}\right)_n & = & \hat m_c \cdot \vec \tau_\beta \nonumber \\
\left( \frac{\partial y}{\partial e}\right)_n & = & \hat t_c \cdot \vec \tau_\beta \nonumber \\
\left( \frac{\partial x}{\partial n}\right)_e & = & \hat m_c \cdot \vec \tau_\alpha \nonumber \\
\left( \frac{\partial y}{\partial n}\right)_e & = & \hat t_c \cdot \vec \tau_\alpha \>, 
\ea
such that we can approximate $\vec \tau_\beta \approx \vec \tau_{\beta_c}$ and  $\vec \tau_\alpha \approx \vec \tau_{\alpha_c}$ near the critical point when needed. Also, near the critical point, one has the useful relationships,
\be
x \approx \Delta e \> \hat m_c \cdot \vec \tau_{\beta_c} + \Delta n \> \hat m_c \cdot \vec \tau_{\alpha_c} \label{xaprox}
\ee
and
\be
y \approx \Delta e \> \hat t_c \cdot \vec \tau_{\beta_c} + \Delta n \> \hat t_c \cdot \vec \tau_{\alpha_c} \>.\label{yaprox}
\ee
We are then left with calculating derivatives of $z$ with respect to $x$ and $y$ in Eqs. (\ref{betaxy}) and (\ref{alfaxy}).\\

{\bf The order parameter.} For this, we use Eq. (\ref{betaxy}) for the temperature $\beta$, and the conditions that at coexistence $x \approx - c_0 y^{2\Delta}$ and the derivative of $z$ with respect to $y$ is zero. Using the scaling form given by Eq. (\ref{scal}) and the above equations, one finds initially,
\begin{equation}
\beta \approx \beta_c - \sqrt{1+\alpha_c^2+\beta_c^2} \> y^{2\Gamma_0 - 2 \Delta} {\cal F}^\prime(-c_0) \> \left( \frac{\partial x}{\partial e}\right)_n\>,\label{betacoex}
\end{equation}
and,
\begin{equation}
\Delta e \approx - \frac{\hat m_c \cdot \vec \tau_{\alpha_c}}{\hat m_c \cdot \vec \tau_{\beta_c}} \Delta n \>.
\end{equation}
This leads to an expression of $y$ in terms of $\Delta n$ at coexistence near the critical point,
\begin{equation}
y \approx \hat t_c \cdot \left[ \vec \tau_{\alpha_c} -   \vec \tau_{\beta_c} \frac{\hat m_c \cdot \vec \tau_{\alpha_c}}{\hat m_c \cdot \vec \tau_{\beta_c}} \right] \Delta n \>.
\end{equation}
Substituting this into Eq. (\ref{betacoex}) yields the desired relationship between the temperatura $\beta - \beta_c$ and the density $\Delta n$ at coexistence, given by Eq. (\ref{betadn}), from which the critical exponent $\hat \beta$ can be read off. \\

{\bf The critical isotherm.} Here we search for the relationship between the chemical potential $\alpha$ and the density $\Delta n$, as the critical point is approached, along the critical isotherm $\beta = \beta_c$. As indicated in the main text, we start once more with Eq.(\ref{betaxy}), set $\beta = \beta_c$ and calculate the derivatives using the scaling form of $z(x,y)$, Eq. (\ref{scal}). This gives
\begin{equation}
0 \approx \vec \tau_{\beta_c} \cdot \left[\hat m_c \> y^{2\Gamma-2\Delta} {\cal F}^\prime\left(\frac{x}{y^{2\Delta}}\right) +
\hat t_c  \> y^{2\Gamma_0-1} \left(2\Gamma_0 {\cal F}\left(\frac{x}{y^{2\Delta}}\right) - 2\Delta \frac{x}{y^{2\Delta}}{\cal F}^\prime\left(\frac{x}{y^{2\Delta}}\right)\right)\right]\>.\label{zero}
\end{equation}
There is a very interesting issue here. The solution to this equation is not a single point but a curve on the entropy surface that should be expressed as $x = x(y)$. Hence, the knowledge of such a function should allows us to write down the previous equation as, in general, a  non-analytic power series in $y$, indicating then that the coefficients of such a power series must vanish separately. By observing (\ref{zero}) we find that the lowest order in $y$ is $y^{2\Gamma-2\Delta}$, since $2\Gamma_0 - 2 \Delta < 2\Gamma_0 -1$, for $\Delta \ge 1$. This indicates that ${\cal F}^\prime$ must be evaluated at a constant value. For this to be true, the critical isotherm must be of the form $x \approx - d_0 y^{2\Delta}$ with $d_0$ a constant coefficient that must obey $d_0 < c_0$, in order for the critical isotherm to be within the surface. This implies then that
\begin{equation}
{\cal F}^\prime(-d_0) = 0 \>.\label{condisot1}
\end{equation}
Now we face an apparent difficulty in Eq. (\ref{zero}). It seems to indicate, following the same reasoning, that the coefficient multiplying the power $y^{2\Gamma_0-1}$ must also vanish at the critical isotherm, implying that ${\cal F}(-d_0)$ should be zero. But this cannot be true since ${\cal F}(X)$ is different from zero for all  values of $X$. Hence, in order for the coefficient of $y^{2\Gamma_0-1}$ to become zero, there must be a non-vanishing correction term arising from the derivative $(\partial z/\partial x)_y$, that cancels the leading term in $(\partial z/\partial y)_x$. That is, Eq.(\ref{zero}) must be corrected in the following fashion,
\begin{equation}
0 \approx \vec \tau_{\beta_c}\cdot \left[\hat m_c \> \left(y^{2\Gamma-2\Delta} {\cal F}^\prime(-d_0) + {\cal D} y^{2\Gamma_0-1}\right)
\hat t_c \>  y^{2\Gamma_0-1} \left(2\Gamma_0 {\cal F}(-d_0) + 2\Delta d_0{\cal F}^\prime(-d_0)\right)\right]\>.\label{correc2}
\end{equation}
such that Eq. (\ref{condisot1}) holds to lowest order and, to the next $y^{2\Gamma_0-1}$, one obtains,
\begin{equation}
0 \approx \vec \tau_{\beta_c} \cdot \left[\hat m_c \> {\cal D} +2 \Gamma_0
 \hat t_c \> {\cal F}(-d_0) \right]\>,\label{correc3}
\end{equation}
from which we identify the value of ${\cal D}$, valid only at the critical isotherm, in terms of critical properties and the scaling function.  The above results can now be substituted into the expression for the chemical potential $\alpha$, given by Eq.(\ref{alfaxy}), leading directly to Eq. (\ref{alfay}), from which the critical exponent $\hat \delta$ is obtained.\\

{\bf The specific heat at constant volume.} For this derivation and the following for the isothermal compressibility, we will use the scaling form ${\cal G}$, Eq. (\ref{G2}), valid for $x >0$, allowing us to set $\Delta n = 0$ and let $T \to T_c$ from above. As mentioned in the text, this the simplest way to study the critical behavior of $c_v$ and $\kappa_T$. We start by writing the specific heat as,
\begin{equation}
-\beta^2 c_v^{-1} =  \frac{\partial \beta}{\partial e} \>,\label{cn2}
\end{equation}
and use again the expression for $\beta$ given by Eq. (\ref{betaxy}). For the sake of clarity and because we will use them for the calculation of $\kappa_T$, we first write the exact expression for the derivative of $\beta$ with respect to $e$,
\ba
 \frac{\partial \beta}{\partial e} & =  \sqrt{1+\alpha_c^2 + \beta_c^2}&\left[ \left(\frac{\partial^2 z}{\partial x^2}\right) \left( \hat m \cdot \vec \tau_\beta\right)^2 +2\left(\frac{\partial^2 z}{\partial x\partial y}\right)  \left(\hat m \cdot \vec \tau_\beta \right)\left(\hat t \cdot \vec \tau_\beta \right)+ \left(\frac{\partial^2 z}{\partial y^2}\right)  \left(\hat t \cdot \vec \tau_\beta \right)^2 \right. \nonumber \\
&&\left. +\left(\frac{\partial z}{\partial x}\right)\hat m \cdot (0,0,\frac{\partial \beta}{\partial e} )+ \left(\frac{\partial z}{\partial y}\right) \hat t \cdot (0,0,\frac{\partial \beta}{\partial e})\right] \>,\label{cvmed}
\ea
and their corresponding values in terms of the scaling function ${\cal G}$,
\be
\frac{\partial z}{\partial x} = - \frac{\Gamma_0}{\Delta} x^{\frac{\Gamma_0}{\Delta} -1} {\cal G}\left(\frac{y^2}{x^{1/\Delta}}\right) + \frac{1}{\Delta}  x^{\frac{\Gamma_0}{\Delta} -1} \>\frac{y^2}{x^{1/\Delta}}{\cal G}^\prime\left(\frac{y^2}{x^{1/\Delta}}\right) \label{dzx}
\ee
\be
\frac{\partial z}{\partial y} = - 2 x^{\frac{\Gamma_0}{\Delta} -\frac{1}{\Delta}} \> y \> {\cal G}^\prime \left(\frac{y^2}{x^{1/\Delta}}\right) \label{dzy}
\ee
\ba
\frac{\partial^2 z}{\partial x^2} & = & - \frac{\Gamma_0}{\Delta}\left(\frac{\Gamma_0}{\Delta}-1\right) x^{\frac{\Gamma_0}{\Delta} -2} {\cal G}\left(\frac{y^2}{x^{1/\Delta}}\right) 
+\left[\frac{\Gamma_0}{\Delta^2} + \frac{1}{\Delta}\left(\frac{\Gamma_0}{\Delta}-1\right) - \frac{1}{\Delta^2}\right] x^{\frac{\Gamma_0}{\Delta} -2}\>\frac{y^2}{x^{1/\Delta}} {\cal G}^\prime\left(\frac{y^2}{x^{1/\Delta}}\right) \nonumber \\
&&- \frac{1}{\Delta^2} x^{\frac{\Gamma_0}{\Delta} -2}\> \left(\frac{y^2}{x^{1/\Delta}}\right)^2 {\cal G}^{\prime\prime}\left(\frac{y^2}{x^{1/\Delta}}\right) \label{d2zxx}
\ea
\ba
\frac{\partial^2 z}{\partial x\partial y} & = &-2\left(\frac{\Gamma_0}{\Delta}-\frac{1}{\Delta}\right)\>y\>x^{\frac{\Gamma_0}{\Delta}-\frac{1}{\Delta}-1}{\cal G}^\prime\left(\frac{y^2}{x^{1/\Delta}}\right)+\frac{2}{\Delta}\>y\>x^{\frac{\Gamma_0}{\Delta}-\frac{1}{\Delta}-1}\>\frac{y^2}{x^{1/\Delta}}{\cal G}^{\prime\prime}\left(\frac{y^2}{x^{1/\Delta}}\right) \label{d2zxy}
\ea
\ba
\frac{\partial^2 z}{\partial y^2} & = & -2\>x^{\frac{\Gamma_0}{\Delta}-\frac{1}{\Delta}}{\cal G}^\prime\left(\frac{y^2}{x^{1/\Delta}}\right)-4 \>x^{\frac{\Gamma_0}{\Delta}-\frac{1}{\Delta}}\>\frac{y^2}{x^{1/\Delta}}{\cal G}^{\prime\prime}\left(\frac{y^2}{x^{1/\Delta}}\right)\>.\label{d2zyy}
\ea
These expressions are valid for all values of $y$ and $x > 0$. Now, our goal is to find $c_v$ to the lowest leading order when $\Delta n = 0$. From the above expressions, one concludes that this should be given in terms of powers of $x$. For this we recall the relationships between $x$ and $y$ in terms of $\Delta e$ and $\Delta n$, near the critical point, Eqs. (\ref{xaprox}) and (\ref{yaprox}), and find that at the line $\Delta n = 0$, $x$ and $y$ are linearly related by
\be
y \approx \frac{\hat t_c \cdot \vec \tau_{\beta_c}}{\hat m_c \cdot \vec \tau_{\beta_c}} x \>.\label{yx}
\ee
Using this expression into $c_v$ given by Eq. (\ref{cvmed}) and the approximated forms Eqs (\ref{dzx})-(\ref{d2zyy}), we can find an expression in powers of $x$. In particular, since ${\cal G}(Y)$ is an analytic power series in terms of its argument $Y$, with ${\cal G}(0) > 0$, one can set $Y = 0$ in ${\cal G}$, ${\cal G}^\prime$ and ${\cal G}^{\prime \prime}$ and treat them as constant. Inspecting the ensuing powers in $x$ one finds that the smallest exponent is $\frac{\Gamma_0}{\Delta} - 2$. This yields $c_v$ depending on $x$ for $\Delta n = 0$,
\be
\beta^2 c_v^{-1} \approx \sqrt{1+\alpha_c^2 + \beta_c^2}\>\hat m_c \cdot \vec \tau_{\beta_c}\>\frac{\Gamma_0}{\Delta}\left(\frac{\Gamma_0}{\Delta}-1\right) x^{\frac{\Gamma_0}{\Delta} -2}\>{\cal G}\left(0\right) \>.\label{cv1}
\ee
Since we want to find the temperature dependence of $c_v$, we now write the $\beta$ near the critical point as a function of $x$ and $y$ using Eq. (\ref{betaxy}) 
\begin{equation}
\beta \approx \beta_c + \sqrt{1+\alpha_c^2+\beta_c^2} \left[\left(\frac{\partial z}{\partial x}\right)_y\hat m_c \cdot \vec \tau_{\beta_c} + \left(\frac{\partial z}{\partial y}\right)_x \hat t_c \cdot \vec \tau_{\beta_c} \right]\>.
\end{equation}
Using the expressions given by Eqs. (\ref{dzx}) and (\ref{dzy}) for the derivatives, and proceeding as above with $c_v$, one finds a power series in $x$, whose leading order is,
\be
\beta \approx \beta_c - \sqrt{1+\alpha_c^2 + \beta_c^2} \>\hat m_c \cdot \vec \tau_{\beta_c} \frac{\Gamma_0}{\Delta} x ^{\frac{\Gamma_0}{\Delta}-1} {\cal G}(0)  \>.\label{betay0}
\ee
Combination of Eqs. (\ref{cv1}) and (\ref{betay0}) yields the specific heat at constant volumen, for $\Delta n = 0$, as $T\to T_c$ from above, as given by Eq. (\ref{cnfinal}). The $\hat \alpha$ critical exponent follows.\\

{\bf Isothermal compressibility.} To perform this calculation is useful to rewrite the isothermal compressibility as,
\begin{eqnarray}
-\frac{\beta}{n^2} \kappa_T^{-1} & = & \frac{\partial^2 s}{\partial n^2}  - \frac{\left(\frac{\partial^2 s}{\partial e\partial n}\right)^2 }{\frac{\partial^2 s}{\partial e^2}} \nonumber \\
& = & - \frac{\partial \alpha}{\partial n} - \frac{\left(\frac{\partial \beta}{\partial n}\right)^2 }{\frac{\partial \beta}{\partial e}}\>,\label{kt}
\end{eqnarray}
where it is understood that $\beta = \beta(e,n)$ and $\alpha = \alpha(e,n)$. For completeness, we write the derivatives of $\alpha$ and $\beta$ with respect to $n$:
\ba
\frac{\partial \alpha}{\partial n}  =&  -\sqrt{1+\alpha_c^2 + \beta_c^2}&\left[ \left(\frac{\partial^2 z}{\partial x^2}\right) \left( \hat m \cdot \vec \tau_\alpha\right)^2 +2\left(\frac{\partial^2 z}{\partial x\partial y}\right)  \left(\hat m \cdot \vec \tau_\alpha \right)\left(\hat t \cdot \vec \tau_\alpha \right)+ \left(\frac{\partial^2 z}{\partial y^2}\right)  \left(\hat t \cdot \vec \tau_\alpha \right)^2 \right. \nonumber \\
&&\left. +\left(\frac{\partial z}{\partial x}\right)\hat m \cdot (0,-\frac{\partial \alpha}{\partial n} ,0 )+ \left(\frac{\partial z}{\partial y}\right) \hat t \cdot (0,-\frac{\partial \alpha}{\partial n} ,0 )\right] \>.
\ea
\ba
\frac{\partial \beta}{\partial n}  =& \sqrt{1+\alpha_c^2 + \beta_c^2}&\left[ \left(\frac{\partial^2 z}{\partial x^2}\right) \left( \hat m \cdot \vec \tau_\beta\right)\left( \hat m \cdot \vec \tau_\alpha\right) +\left(\frac{\partial^2 z}{\partial x\partial y}\right) \left[ \left(\hat m \cdot \vec \tau_\beta \right)\left(\hat t \cdot \vec \tau_\alpha \right)+ \left(\hat m \cdot \vec \tau_\alpha \right)\left(\hat t \cdot \vec \tau_\beta \right)\right] \right. \nonumber  \\
&&+ \left(\frac{\partial^2 z}{\partial y^2}\right)  \left(\hat t \cdot \vec \tau_\beta \right) \left(\hat t \cdot \vec \tau_\alpha \right)  \left. +\left(\frac{\partial z}{\partial x}\right)\hat m \cdot (0,0,\frac{\partial \beta}{\partial n} )+ \left(\frac{\partial z}{\partial y}\right) \hat t \cdot (0,0,\frac{\partial \beta}{\partial n})\right] \>.
\ea
Once again, we want to find $\kappa_T$ for $\Delta n  = 0$, as $T \to T_c$ from above. We follow the same procedure as in the case of the specific heat, using the linear relationship between $x$ and $y$, Eq. (\ref{yx}), and write all terms in terms of powers of $x$. Here, we must be very careful since the leading order term in the expression for $\kappa_T$, Eq. (\ref{kt}), vanishes! Hence, we must go to the next order. It is a very lengthy but straightforward calculation that leads to an expression of $\kappa_T$ with a leading order $\sim x^{\frac{\Gamma_0-1}{\Delta}}$ which, with the use of Eq. (\ref{betay0}), gives the desired expression for the iosthermal compressibility given by Eq. (\ref{kappafinal}). Then we obtain the critical exponent $\hat \gamma$.

\end{document}